# Atomic Pathways of Solute Segregation in the Vicinity of Nanoscale Defects

Samik Mukherjee,[†, *] Simone Assali,[†] and Oussama Moutanabbir[†, *]

[†]*Department of Engineering Physics, École Polytechnique de Montréal, C. P. 6079, Succ. Centre-Ville, Montreal, Québec H3C 3A7, Canada*
*Email: samik.mukherjee@polymtl.ca; oussama.moutanabbir@polymtl.ca

**Abstract:**

This work unravels the atomic details of the interaction of solute atoms with nanoscale crystalline defects. The complexity of this phenomenon is elucidated through detailed atom probe tomographic investigations on epitaxially-strained, compositionally metastable, semiconductor alloys. Subtle variations are uncovered in the concentration and distribution of solute atoms surrounding dislocations, and their dynamic evolution is highlighted. The results demonstrate that crystal defects, such as dislocations, are instrumental in initiating the process of phase separation in strained metastable layers. Matrix regions, close to the dislocations, show clear signs of compositional degradation only after a relatively short time from disrupting the local equilibrium. The solute concentration as well as the density of non-random atomic clusters increases while approaching a dislocation from the surrounding matrix region. In parallel, far from a dislocation the lattice remains intact preserving the metastable structure and composition uniformity. At advanced stages of phase separation, the matrix outside the dislocation reaches the equilibrium concentration, while dislocations act as vehicles of mass-transport, providing fast diffusive channels for solute atoms to reach the surface. This process occurs by the steady increase in the solute concentration rate of ~0.5 at.%. per 10 nm of the dislocation. Besides, the number of atomic clusters almost doubles and the number of atoms per cluster increases steadily, moving along a dislocation towards the surface. In addition to understanding the atomic features involved in the phase separation of strained metastable alloys, the work also illustrates the thermodynamic and kinetic behavior of solute atoms in the vicinity of a nanoscale defect and describe quantitatively the key processes, thus providing the empirical input to improve the atomic models and simulations.

The behavior of a solute in a lattice is affected by its interaction with crystal defects such as dislocations,[1,2] stacking faults,[3] twins, and grain boundaries.[4,5] These interactions can shape key material properties including boundary strength, yield strength, oxidation, and precipitate growth kinetics.[3,6–8] Solute-defect interactions are usually driven by electrostatic effects, size misfit effects, and modulus effects.[9] The former rises since both crystal defects and solute atoms may have electrical charges (especially in intrinsic semiconductors and dielectrics where there are no mobile charges to screen the field), while the latter two are strain-related, originating from the difference in size and the elastic moduli, between the solute and the host material, respectively. Additionally, depending on the boundary conditions, the solute-dislocation interaction manifests itself within materials differently. It was suggested that the compositional variations induced by atomic segregation at the surface, and the associated surface diffusion and trapping by atoms, act as the driving force for the nucleation and migration of dislocations at the sub-surface regions of a multi-component alloy.[10] Others report that dislocations that nucleate first in strained epitaxial layers, thus providing the driving force to solute atoms to nucleate into aggregate phases.[1–3]

An epitaxially-grown strained metastable alloy is an even more complex playground, wherein a combination of epitaxial strain, compositional metastability, and defect strain provide additional driving forces, like the breakdown of bulk thermodynamics and suppression of precipitation barrier, to influence the interaction of the solute atoms with crystal defects. Elucidating the precise behavior of the solute atoms in these material systems is crucial to understand at the fundamental level their structural stability in presence of several interrelated processes, and eventually control the material properties. However, such crystal imperfections are inherently three-dimensional (3D) in nature and are often only a few atomic layers thick, making it a formidable challenge to accurately probe the behavior of solute atoms in their vicinity.[11] Atom probe tomography (APT) is one of the outstanding techniques that can be used to study the local, atomic-level chemical environments at crystal defects,[12–14] thereby presenting a 3D view of the solute-defect interaction. With this perspective, this work employs APT to track atomic pathways defining the behavior of solute atoms at nanoscale crystal defects in epitaxially-strained, compositionally metastable, semiconductor layers. Using epitaxial germanium-tin (GeSn) as a model system, the results and their associated discussions in the following sections elucidate the precise role played by the crystal imperfections at various stages of the phase separation process in strained metastable materials.

The growth of α-Sn saturated GeSn epitaxial layers, on Ge buffered silicon (Si) substrates, was carried out in a low-pressure chemical vapor deposition (CVD) chamber, using monogermane and tin-chloride as precursors, and hydrogen ($H_2$) as the carrier gas.[15] The GeSn stack consists of three layers with a top-layer (TL), a middle layer (ML), and a bottom-layer (BL), having a thickness (Sn concentration) of ~160 nm (~18.0 at.%), ~155 nm (~11.0-14.0 at.%), and ~65 nm (~8.0 at.%), respectively. These layers are metastable as the equilibrium solid solubility of Sn in Ge at room temperature is less than 1.0 at. %. The residual strain in TL, ML, and BL layers is ~ −1.27 %, ~ −0.54 %, and ~ −0.20 %, respectively. Post-growth, the samples were annealed in a $H_2$-rich atmosphere at 320 ºC to promote solute diffusion and their segregation and investigate the atomistic behavior during the phase separation. See the STEM image of the as-grown and the annealed samples in Figure S1 of the Supplementary Information. Figure 1(a) shows a high-resolution STEM image of a dislocation and the associated electron energy loss spectroscopy (EELS) map. The map shows the enrichment (dilution) of Sn (Ge) atoms at the dislocation.

However, the EELS signal being averaged over the entire STEM lamella thickness and fails to reveal any subtle variation in the 3D distribution of Sn.

Next, an atom-by-atom mapping was conducted by APT using a LEAP5000 system, in the pulsed-laser mode (see details in Supplementary Information). To improve the odds of mapping a dislocation, multiple APT nanotips were fabricated from the samples, using the focussed ion-beam lift-out and sharpening technique, described elsewhere.[16] Figure 1(b) shows the atom-by-atom map of Sn atoms after 30 min-annealing. Also overlaid in the map are the Sn iso-concentration surfaces, drawn at 1.75 at.% Sn concentration. The dislocation, identified by the columnar pipe-like Sn-segregated region in the APT map (see double black arrows), is demarcated from the rest of the matrix by the iso-concentration surface. From here onwards in the text, the atomic column formed due to the Sn segregation at the dislocation will be called the 'Sn-pipe'. At the bottom of the Sn-pipe lies the GeSn(BL)/Ge interface. Figure 1(b) also reveals that some Sn enrichment took place at the BL/Ge interface. Such enrichments occur due to the lower energy barrier provided to Sn atoms by pre-existing defective interfaces and dislocation lines, to nucleate into aggregate phases (clusters and precipitates). The probability of aggregate formation increases in the vicinity of crystal defects and interfaces when the local strain around them lowers the aggregate-alloy interface energy.[9] To evaluate any subtle variations in the Sn concentration in the vicinity of the dislocation, two coplanar cylinders (10 nm diameter) were placed in the orientation depicted in the inset of Figure 1(b). The Sn concentration profiles along the length of these cylinders are shown in Figure 1(c). The obtained profiles are similar, reaching a maximum concentration of ~3.4 at.% within the Sn-pipe, while the matrix is close to the equilibrium concentration (~0.8 at.% of Sn in Ge).

The distribution of atoms near the Sn-pipe was also evaluated. A cube labeled C1' was defined close to the dislocation, as shown in Figure 1(b). Atoms located within C1' were extracted for subsequent statistical analysis, namely the $k^{th}$ order nearest-neighbor distribution $(NND_k)$[17] and the partial radial distribution (p-RDF)[18] functions. Figure 1(d) displays the $NND_k$ (for k =1 and 2) of the Sn atoms located within C1'. The distributions obtained from the APT data are shown as histograms, while that of the corresponding randomized data set are shown using solid black lines. A randomized data set is one wherein the chemical identity of each atom within the 3D distribution is randomized while keeping their position unchanged. The analysis shows that even at the immediate vicinity of the dislocation, Sn atoms are randomly distributed within the GeSn matrix. The randomness was further verified by the Sn p-RDF analysis (relative to Sn), as shown in the inset of Figure 1(d). A p-RDF evaluates the correlations (self-correlation, as in this case, between the same atomic species, or cross-correlation between different atomic species) in the atomic distribution within a region of interest.[18,19] The p-RDF in the inset of Figure 1(d) can be seen to steady out at the value of unity, indicating the lack of any positive or negative correlation between Sn solute atoms within C1'. A similar analysis was carried out by extracting atoms from the TL of in different APT data set of the same GeSn sample (annealed at 320 °C for 30 min), having no dislocation in its vicinity. There too the atoms were found to be randomly distributed (not shown here). The analysis above shows that at an advanced stage of phase separation, except for the Sn-pipes that are formed at the dislocations, the entire GeSn matrix has reached the equilibrium Sn concentration with Sn atoms randomly distributed within the matrix.

To unravel the atomic distribution at earlier stages of segregation and phase separation, APT investigations were also carried out on samples annealed at 320 ℃, but for shorter durations. Figure 2(a) exhibits the 3D APT map recorded for an annealing time of 5 min. Panels P1 and P2 correspond to different regions of the sample, from two different APT data sets. Iso-concentration surfaces, drawn at different Sn concentrations are overlaid in the 3D maps in each panel. P1 shows the atomic map, along with a Sn-pipe (double black arrows), formed due to atomic segregation at a dislocation. The APT map in panel P2 is clipped around the region of the BL/Ge interface. The map does not contain dislocation and no Sn segregation is seen at the interface. Compared to the typical length scale of an APT data set, this region can be considered to be far from any dislocation. This cube C3 within the atomic map in panel P2 is used as a reference for comparison during the subsequent $NND_k$ analysis. Figure S2 in the Supplementary Information is a third APT data set of the same sample, showing a solidified Sn droplet at the surface. The Sn droplet is formed due to the migration of Sn atoms from subsurface layers and their subsequent segregation at the surface.

The proximity histogram depicting the variation of Sn concentration within and outside the solidified Sn droplet is also shown in Figure S2. A proximity histogram is a weighted superposition of concentration profiles obtained by binning atoms in 3D, at fixed incremental distances from the iso-concentration surface. The histogram reveals that, on approaching the droplet from the surrounding matrix, the Sn concentration increases, first gradually and then abruptly until finally, it reaches the equilibrium content of ∼99 at.% within the droplet, while far from the droplet the mean Sn concentration within the TL reduced from the initial value of ∼18 at.% to ∼14.4 at.%. Figure S3 in the Supplementary Information shows the mass-spectrum of a fourth data set. The latter was recorded from the TL (atomic distribution is shown in the inset) of the same sample. Figure S3 indicate that when there is no solidified Sn droplet at the surface, the TL remains intact.

The mass-spectrum analysis found the mean Sn concentration within the TL of the data set to be ~17.7 at.% similar to the as-grown concentration (~18.0 at.%), with no disruption in their atomic distribution.

Figure 2(b) shows the Sn concentration profile along the length of the three coplanar cylinders that intersect the Sn-pipe in panel P1. The orientation of the cylinders, relative to the Sn-pipe is shown in the inset. A clear difference is observed when compared to the profile shown in Figure 1(c). On approaching the dislocation, the Sn concentration along all three cylinders can be seen to increase. At the dislocation, the Sn concentration reaches up to ~14 at.%. The gradual increase in Sn content is due to the migration of Sn atoms from the surrounding matrix towards the dislocation. Figure 2(b) also reveals that there exists a spatial variation of the Sn concentration surrounding the dislocation. First, the rate at which the Sn concentration increases varies depending on the cylinder orientation. Second, close to the dislocation, the Sn concentration along the black cylinder takes a sudden dip, while on the horizontally opposite side, Sn concentration shows a bump. The magnitude of the dip and the bump in Sn concentration can also be seen to change when the cylinder orientation changes. Such behavior is indicative of a nonuniformity in the local strain field around the dislocation, which drives Sn atoms from the compressive side to the tensile side.[20,21]

Figure 3 shows the $NND_k$ analysis (for k = 1,2) for the Sn atoms within the cubes C1, C2, and C3. See the positioning of the three cubes in Figure 2(a). C1 is closest to the Sn-pipe while C2 is a little further out (see panel P1 in Figure 2(a)). C3 is placed within the GeSn BL, far from any dislocation (see panel P2 in Figure 2(a)). The bottommost panel in Figure 3 shows that atoms within C3 are randomly distributed, as revealed by the good overlap of the $NND_k$ data, obtained from APT, with the corresponding randomized distribution. The randomness of atomic distribution in C3 was additionally verified by performing p-RDF and frequency distribution analysis, as shown in Figure S4 of the Supplementary Information. C2 (middle panel) shows a bimodal distribution with a smaller satellite peak to the left of the stronger main peak. The identities of the two peaks are clearly distinguished by the Gaussian fitting curves, shown using the solid green and red lines. The bimodal $NND_k$ data sets indicate that the Sn distribution within C2 deviates from the randomized distribution. The weaker satellite peak represents little pockets within C2, wherein the average nearest-neighbor distances are smaller than the rest of the matrix. This hints to the presence of non-random atomic clusters (NRAC). The presence of NRAC is confirmed by performing the cluster distribution analysis, based on the maximum-separation algorithm (details given elsewhere[22,23]). The 3D distribution of the NRAC within C2 is also shown. Once the clusters are identified, it is easy to separate the NRAC from the remaining matrix to subsequently perform the $NND_k$ analysis. When done separately, the $NND_k$ analysis of the GeSn matrix and the NRAC are both unimodal (not shown here). $NND_1$ and $NND_2$ of the GeSn matrix were found to show maxima at ~0.27 nm and ~0.43 nm, respectively. Taking into account the 80% detector efficiency, these values correspond to $1^{st}$ and $2^{nd}$ nearest-neighbor distances of ~0.24 nm and ~0.40 nm, respectively, close to the values expected for diamond cubic GeSn layers.[24] Similarly, the $1^{st}$ and $2^{nd}$ nearest-neighbor distances of the NRAC are at ~0.20 nm and ~0.30 nm, respectively.

A similar analysis is shown for cube C1, which is closer to the dislocation as compared to C2. The satellite peak on the left can be seen to increase in magnitude, for both $NND_1$ and $NND_2$, indicating a larger contribution made by the NRAC to the total $NND_k$. Indeed, this is verified from the cluster analysis (displayed alongside), showing a higher density (~ factor 3) of NRAC in C1, as compared to C2. A pile-up could have occurred as the atoms from the surrounding GeSn matrix started to migrate towards the dislocation on receiving the thermal energy. This is most likely what led to the formation of the NRAC. Note, C1 and C1' are equivalent, as far as the position of the cubes relative to the dislocation is concerned. Yet, there is a contrasting difference in the distribution of Sn atoms within the two cubes – a strong deviation from randomness with a high density of NRAC in C1 and a random distribution with no NRAC within C1'. A separate analysis revealed (not shown here) that the most probable number of solute atoms per NRAC within C1 and C2 is 12-15 and 7-9, respectively. Since 20% of all atoms go undetected in APT, the actual number of Sn atoms per NRAC could be higher by a factor of 1.25. The analysis above paints an atomic-level picture of Sn distribution inside GeSn, in the early stages of solute-dislocation interaction. Far from any dislocations, atoms within the GeSn matrix remain randomly distributed with no evidence of any formation of NRAC. Near a dislocation, the scenario changes. The Sn concentration starts to increase steadily closer to a dislocation. Alongside, the atomic distribution is no longer random, with the deviation from randomness intensifying, and the density as well as the size (average number of atoms per NRAC) increasing, close to a dislocation.

Next, the atoms within the Sn-pipes were extracted out for analysis. The top panels in Figure 4 refer to the Sn concentrations, while the bottom panels refer to the cluster statistics within the Sn-pipes found within the samples annealed for 5 min (Figure 4(a)) and 30 min (Figure 4(b)), respectively. Looking at the two concentration profiles in the top panels, a clear difference is visible. While the Sn concentration increases steadily at the rate of ~0.5 at.%. per 10 nm, within the Sn-pipe in Figure 4(b), it can be seen to remain almost constant, all along the length of the Sn-pipe in Figure 4(a). Next, the Sn-pipes were divided into three roughly equal sections, as shown by the black dotted lines in Figure 4. From cluster analysis conducted within each section of the Sn-pipes (Figure 4, bottom), the number of atoms per NRAC was found to have a large distribution and was therefore categorized into three classes: i) 4-8 atoms, ii) 9-14 atoms, and iii) >14 atoms. The lower panels in Figure 4 show that, at an early stage of solute-dislocation interaction, the total number of NRAC remains almost the same while moving from the bottom to the top of the Sn-pipe, where 4-8 atom NRACs are predominating. Moreover, the number of large NRACs (having more than 14 atoms) remains low throughout the length of the Sn-pipe. At an advanced stage of solute-dislocation interaction, the total number of NRAC increases steadily from the bottom to the top of the Sn-pipe. Additionally, while the lowermost section of the Sn-pipe is dominated by 4-8 atom NRAC, the middle and the topmost sections are dominated by larger NRAC having 9 atoms or more. The above analysis highlights the evolution in the solute-dislocation interaction as the solute diffuses towards the surface during annealing. At an early stage, the Sn concentration and the number and size of the NRAC remain constant along the length of the dislocation, indicating that only a limited number of solutes diffuse through the dislocations towards the surface. For extended annealing, the Sn concentration increases steadily along the length of the dislocation and so do the NRAC number and size, indicating the presence of a steady upward diffusion of the solute atoms along the dislocation.

The relaxation process of metastable alloys upon thermal treatment is a strong function of the initial state of the material. For example, fully-strained pseudomorphic metastable GeSn alloys are found to relax the epitaxial strain upon thermal annealing, primarily by Sn diffusion and subsequent surface segregation before the onset of plastic relaxation.[25] The critical temperature, marking the onset of sudden Sn diffusion, could be significantly higher than the typical GeSn growth temperatures used in CVD, especially for low (≤ 8 at.%) Sn containing layers. The difference between this critical temperature and the growth temperature vanishes progressively as the layer's Sn content increases. In partially relaxed layers with pre-existing dislocations, the scenario is quite different, with the layers initially showing further plastic relaxation and the onset of Sn diffusion taking place at much lower temperatures,[26] as it was observed in the highly dislocated regions of the current samples. The Sn atoms in GeSn have a strong tendency to reach the surface from the bulk layers. At the surface, the atoms can diffuse freely with a nominal diffusion barrier. This allows the atoms to easily form β-Sn clusters, which is the thermodynamically stable phase at room temperature. Surface segregation of Sn was evidenced by several experimental investigations on GeSn[27–30]. Vacancy-assisted diffusion might play a key role in the diffusion at a smaller length, such as to the surface from the sub-surface layers or to the dislocations from the neighboring matrix. On the other hand, atomic transport from deeper layers up to the surface is primarily dislocation-assisted, as highlighted in the current work.

While the phase separation in a strained metastable epitaxial layer is always thermodynamically driven, the speed at which the thermodynamic equilibrium is reached is guided by kinetic factors. In this work, the atomic-level mapping carried out on the thermally-annealed strained metastable alloys revealed the following. Evidence of mass-transport of the solute atoms from the surrounding matrix towards the dislocations and subsequent atomic segregation, within the highly-dislocated BL of GeSn, was found to occur at a relatively early stage. This manifested itself by a steady increase in the solute concentration and the density of NRAC closer to the dislocation. Meanwhile, the TL (with significantly higher solute content in the as-grown state, compared to the BL) and regions of the BL far off from any dislocation, remained compositionally intact. This clearly reveals that, rather than the absolute solute concentration within the metastable alloy, the presence of crystal defects, in form of incoherent interfaces and dislocations, is instrumental in initiating the process of phase separation. During this initial stage, there is negligible upward transport of atoms along the dislocation. As the process evolves, the dislocations were found to act as vehicles of mass-transport, providing fast diffusive channels to the atoms that segregated at the dislocation cores, to make their way up to the surface, a process known as pipe-diffusion.[31]

In summary, the behavior of solute atoms in the vicinity of a nanoscale defects has been studied in metastable, strained layers. We demonstrated that the interaction of these atoms with crystal defects triggers the process of phase separation in strained metastable alloys. Once initiated, the interaction proceeds through the initial diffusion of some of the excess solute atoms from the matrix surrounding the dislocations, forming atomic clusters in the vicinity and at the core of the dislocations. Subsequently, this process is followed by the gradual diffusion of solute atoms along the dislocation, upward towards the surface, where the atoms can form the Sn-rich phase, thereby restoring the material to its equilibrium concentration and completing the phase separation process. In parallel, the number and the average size of atomic clusters were also found to increase when the local equilibrium is disrupted. The matrix outside the dislocation reaches its equilibrium content, and consequently the remaining solute atoms lose the driving force to leave their lattice sites, ceasing any further diffusion towards the dislocation. Moreover, the atomic clusters that formed near dislocations during the initial stage progressively disappear thus proving that they are not stable and can dissociate upon prolonged annealing. Also, because of the extended thermal annealing, nucleating of additional dislocations takes place which could contribute to further relaxation of any residual strain. The framework developed in this work enables tracking the changes in the solute concentration and distribution around crystal defects. Such atomic-level snapshots could serve as detailed inputs to future theoretical investigations of the subtleties associated with the solute-defect interaction process in strained metastable alloys, thus paving the path to more refined defect engineering and modeling of their impact on the stability of a material.

**ASSOCIATED CONTENT**

**Supporting Information.**

The Supporting Information is available free of charge at ACS website.

**Notes**

The authors declare no competing financial interest.

**Acknowledgements**

The authors thank J. Bouchard for the technical support with the CVD system. O.M. acknowledges support from NSERC Canada (Discovery, SPG, and CRD Grants), Canada Research Chairs, Canada Foundation for Innovation, Mitacs, PRIMA Québec, and Defence R&D Canada (Innovation for Defence Excellence and Security, IDEaS).

**References:**

(1) Bonef, B.; Shah, R. D.; Mukherjee, K. Fast Diffusion and Segregation along Threading Dislocations in Semiconductor Heterostructures. *Nano Lett.* **2019**, *19* (3), 1428–1436. https://doi.org/10.1021/acs.nanolett.8b03734.

(2) Chang, L.; Lai, S. K.; Chen, F. R.; Kai, J. J. Observations of Al Segregation around Dislocations in AlGaN. *Appl. Phys. Lett.* **2001**, *79* (7), 928–930. https://doi.org/10.1063/1.1391409.

(3) Titus, M. S.; Rhein, R. K.; Wells, P. B.; Dodge, P. C.; Viswanathan, G. B.; Mills, M. J.; Der Ven, A. Van; Pollock, T. M. Solute Segregation and Deviation from Bulk Thermodynamics at Nanoscale Crystalline Defects. *Sci. Adv.* **2016**, *2* (12). https://doi.org/10.1126/sciadv.1601796.

(4) Nie, J. F.; Zhu, Y. M.; Liu, J. Z.; Fang, X. Y. Periodic Segregation of Solute Atoms in Fully Coherent Twin Boundaries. *Science* **2013**, *340* (6135), 957–960. https://doi.org/10.1126/science.1229369.

(5) Sha, G.; Yao, L.; Liao, X.; Ringer, S. P.; Duan, Z. C.; Langdon, T. G. Segregation of Solute Elements at Grain Boundaries in an Ultrafine Grained Al-Zn-Mg-Cu Alloy. *Ultramicroscopy* **2011**, *111* (6), 500–505. https://doi.org/10.1016/j.ultramic.2010.11.013.

(6) Feng, B.; Yokoi, T.; Kumamoto, A.; Yoshiya, M.; Ikuhara, Y.; Shibata, N. Atomically Ordered Solute Segregation Behaviour in an Oxide Grain Boundary. *Nat. Commun.* **2016**, *7* (1), 1–6. https://doi.org/10.1038/ncomms11079.

(7) Evans, A. G.; Hutchinson, J. W.; Wei, Y. Interface Adhesion: Effects of Plasticity and Segregation. *Acta Mater.* **1999**, *47* (15), 4093–4113. https://doi.org/10.1016/S1359-6454(99)00269-4.

(8) Kobayashi, Y.; Takahashi, J.; Kawakami, K. Effects of Dislocations on the Early Stage of TiC Precipitation Kinetics in Ferritic Steel: A Comparative Study with and without a Pre-Deformation. *Acta Mater.* **2019**, *176*, 145–154. https://doi.org/10.1016/j.actamat.2019.06.055.

(9) Vannarat, S.; Sluiter, M. H. F.; Kawazoe, Y. First-Principles Study of Solute-Dislocation Interaction in Aluminum-Rich Alloys. *Phys. Rev. B - Condens. Matter Mater. Phys.* **2001**, *64* (22), 224203. https://doi.org/10.1103/PhysRevB.64.224203.

(10) Zou, L.; Yang, C.; Lei, Y.; Zakharov, D.; Wiezorek, J. M. K.; Su, D.; Yin, Q.; Li, J.; Liu, Z.; Stach, E. A.; Yang, J. C.; Qi, L.; Wang, G.; Zhou, G. Dislocation Nucleation Facilitated by Atomic Segregation. *Nat. Mater.* **2018**, *17* (1), 56–62. https://doi.org/10.1038/NMAT5034.

(11) Katnagallu, S.; Stephenson, L. T.; Mouton, I.; Freysoldt, C.; Subramanyam, A. P. A.; Jenke, J.; Ladines, A. N.; Neumeier, S.; Hammerschmidt, T.; Drautz, R.; Neugebauer, J.; Vurpillot, F.; Raabe, D.; Gault, B. Imaging Individual Solute Atoms at Crystalline Imperfections in Metals. *New J. Phys.* **2019**, *21* (12), 123020. https://doi.org/10.1088/1367-2630/ab5cc4.

(12) Blavette, D.; Cadel, E.; Fraczkiewicz, A.; Menand, A. Three-Dimensional Atomic-Scale Imaging of Impurity Segregation to Line Defects. *Science* **1999**, *286* (5448), 2317–2319. https://doi.org/10.1126/science.286.5448.2317.

(13) Thompson, K.; Flaitz, P. L.; Ronsheim, P.; Larson, D. J.; Kelly, T. F. Imaging of Arsenic Cottrell Atmospheres around Silicon Defects by Three-Dimensional Atom Probe Tomography. *Science* **2007**, *317* (5843), 1370–1374. https://doi.org/10.1126/science.1145428.

(14) Miller, M. K. Atom Probe Tomography Characterization of Solute Segregation to Dislocations and Interfaces. *J. Mater. Sci.* **2006**, *41* (23), 7808–7813. https://doi.org/10.1007/s10853-006-0518-5.

(15) Assali, S.; Nicolas, J.; Mukherjee, S.; Dijkstra, A.; Moutanabbir, O. Atomically Uniform Sn-Rich GeSn Semiconductors with 3.0–3.5 µm Room-Temperature Optical Emission. *Appl. Phys. Lett.* **2018**, *112* (25), 251903. https://doi.org/10.1063/1.5038644.

(16) Mukherjee, S.; Watanabe, H.; Isheim, D.; Seidman, D. N.; Moutanabbir, O. Laser-Assisted Field Evaporation and Three-Dimensional Atom-by-Atom Mapping of Diamond Isotopic Homojunctions. *Nano Lett.* **2016**, *16* (2), 1335–1344. https://doi.org/10.1021/acs.nanolett.5b04728.

(17) Philippe, T.; De Geuser, F.; Duguay, S.; Lefebvre, W.; Cojocaru-Mirédin, O.; Da Costa, G.; Blavette, D. Clustering and Nearest Neighbour Distances in Atom-Probe Tomography. *Ultramicroscopy* **2009**, *109* (10), 1304–1309. https://doi.org/10.1016/j.ultramic.2009.06.007.

(18) Sudbrack, C.; Noebe, R.; Seidman, D. Direct Observations of Nucleation in a Nondilute Multicomponent Alloy. *Phys. Rev. B* **2006**, *73* (21), 212101. https://doi.org/10.1103/PhysRevB.73.212101.

(19) Mukherjee, S.; Kodali, N.; Isheim, D.; Wirths, S.; Hartmann, J. M.; Buca, D.; Seidman, D. N.; Moutanabbir, O. Short-Range Atomic Ordering in Nonequilibrium Silicon-Germanium-Tin Semiconductors. *Phys. Rev. B* **2017**, *95* (16), 161402. https://doi.org/10.1103/PhysRevB.95.161402.

(20) Xie, H.; Huang, Q.; Bai, J.; Li, S.; Liu, Y.; Feng, J.; Yang, Y.; Pan, H.; Li, H.; Ren, Y.; Qin, G. Nonsymmetrical Segregation of Solutes in Periodic Misfit Dislocations Separated Tilt Grain Boundaries. *Nano Lett.* **2021**, *21* (7), 2870–2875. https://doi.org/10.1021/acs.nanolett.0c05008.

(21) Andersen, D.; Hull, R. Effect of Asymmetric Strain Relaxation on Dislocation Relaxation Processes in Heteroepitaxial Semiconductors. *J. Appl. Phys.* **2017**, *121* (7), 75302. https://doi.org/10.1063/1.4975789.

(22) Dhara, S.; Marceau, R. K. W.; Wood, K.; Dorin, T.; Timokhina, I. B.; Hodgson, P. D. Atom Probe Tomography Data Analysis Procedure for Precipitate and Cluster Identification in a Ti-Mo Steel. *Data Br.* **2018**, *18*, 968–982. https://doi.org/10.1016/j.dib.2018.03.094.

(23) Vaumousse, D.; Cerezo, A.; Warren, P. J. A Procedure for Quantification of Precipitate

Microstructures from Three-Dimensional Atom Probe Data. *Ultramicroscopy* **2003**, *95*, 215–221. https://doi.org/10.1016/S0304-3991(02)00319-4.

(24) Kumar, A.; Demeulemeester, J.; Bogdanowicz, J.; Bran, J.; Melkonyan, D.; Fleischmann, C.; Gencarelli, F.; Shimura, Y.; Wang, W.; Loo, R.; Vandervorst, W. On the Interplay between Relaxation, Defect Formation, and Atomic Sn Distribution in $Ge_{(1-x)}Sn_{(x)}$ Unraveled with Atom Probe Tomography. *J. Appl. Phys.* **2015**, *118* (2), 025302. https://doi.org/10.1063/1.4926473.

(25) Zaumseil, P.; Hou, Y.; Schubert, M. A.; von den Driesch, N.; Stange, D.; Rainko, D.; Virgilio, M.; Buca, D.; Capellini, G. The Thermal Stability of Epitaxial GeSn Layers. *APL Mater.* **2018**, *6* (7), 076108. https://doi.org/10.1063/1.5036728.

(26) Stanchu, H. V.; Kuchuk, A. V.; Mazur, Y. I.; Pandey, K.; de Oliveira, F. M.; Benamara, M.; Teodoro, M. D.; Yu, S.-Q.; Salamo, G. J. Quantitative Correlation Study of Dislocation Generation, Strain Relief, and Sn Outdiffusion in Thermally Annealed GeSn Epilayers. *Cryst. Growth Des.* **2021**, In Press. https://doi.org/10.1021/acs.cgd.0c01525.

(27) Li, H.; Cui, Y. X.; Wu, K. Y.; Tseng, W. K.; Cheng, H. H.; Chen, H. Strain Relaxation and Sn Segregation in GeSn Epilayers under Thermal Treatment. *Appl. Phys. Lett.* **2013**, *102* (25), 251907. https://doi.org/10.1063/1.4812490.

(28) Comrie, C. M.; Mtshali, C. B.; Sechogela, P. T.; Santos, N. M.; van Stiphout, K.; Loo, R.; Vandervorst, W.; Vantomme, A. Interplay between Relaxation and Sn Segregation during Thermal Annealing of GeSn Strained Layers. *J. Appl. Phys.* **2016**, *120* (14), 145303. https://doi.org/10.1063/1.4964692.

(29) Takase, R.; Ishimaru, M.; Uchida, N.; Maeda, T.; Sato, K.; Lieten, R. R.; Locquet, J.-P. Behavior of Sn Atoms in GeSn Thin Films during Thermal Annealing: *Ex-Situ* and *in-Situ* Observations. *J. Appl. Phys.* **2016**, *120* (24), 245304. https://doi.org/10.1063/1.4973121.

(30) Li, H.; Chang, C.; Chen, T. P.; Cheng, H. H.; Shi, Z. W.; Chen, H. Characteristics of Sn Segregation in Ge/GeSn Heterostructures. *Appl. Phys. Lett.* **2014**, *105* (15), 151906. https://doi.org/10.1063/1.4898583.

(31) Legros, M.; Dehm, G.; Arzt, E.; Balk, T. J. Observation of Giant Diffusivity along Dislocation Cores. *Science* **2008**, *319* (5870), 1646–1649. https://doi.org/10.1126/science.1151771.

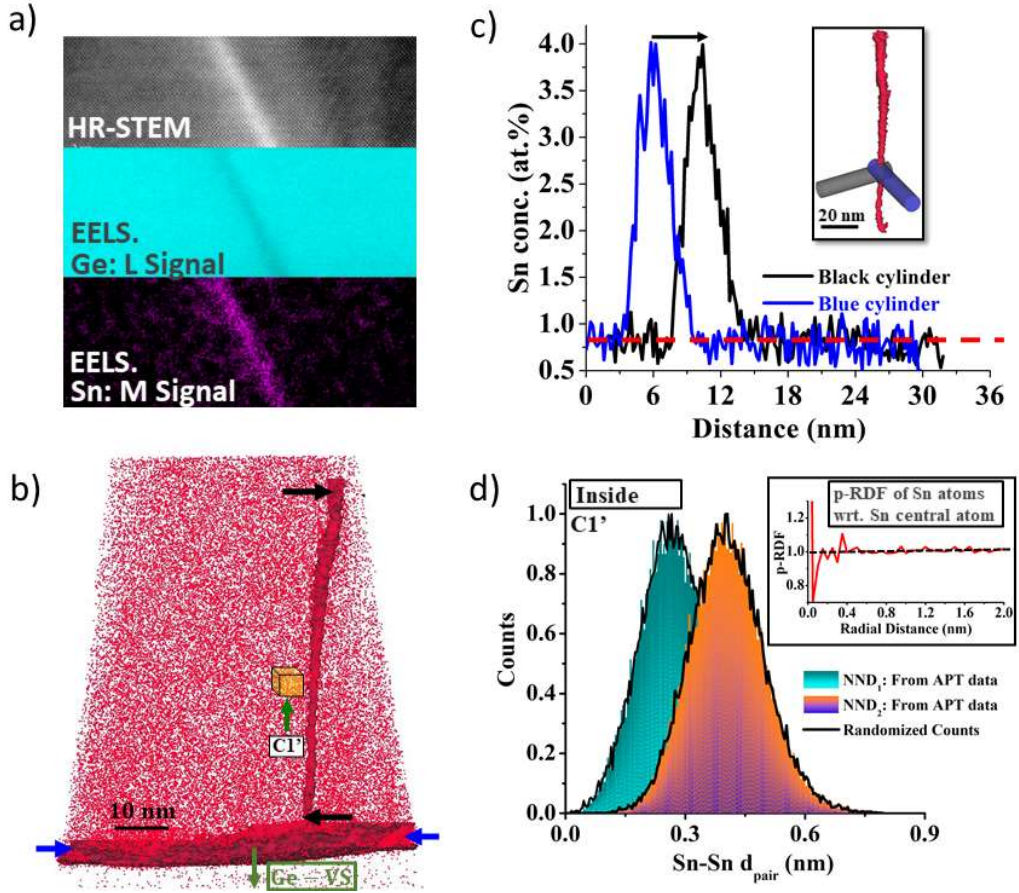

**Figure 1.** (a) High-resolution HAADF-STEM image of dislocation and the corresponding EELS maps of the Ge L signal, and the Sn M signal (inset). The scale bars of the EELS maps is 5 nm. (b) The 3D atom-by-atom map (clipped from the entire reconstruction to highlight the region of interest) of the GeSn sample annealed at 320 °C, for 30 mins, showing a Sn-pipe (marked by the double black arrows) that formed due to segregation of atoms at a dislocation. Only the Sn atoms (red dots) are shown for clarity while the majority element Ge has been left out. The Ge-VS is highlighted, and the interface of the BL with the Ge-VS is marked by the double blue arrows. C1' is a cube, placed proximally to the dislocation. Atoms located within the cube were extracted out for statistical analysis (details in text). Inset: The placement of the cylinders relative to the Sn-pipe. The Sn-pipe was identified by drawing iso-concentration surfaces within the APT map, at 1.75 at.% Sn concentrations. (c) Sn concentration profile along the length of the black and the blue, coplanar cylinders (depicted in the inset of (b)) that intersects the Sn-pipe (marked by the double black arrows in (b)). The profile along the black cylinder is displaced by the length of the black arrow for clarity. The mean Sn concentration within the GeSn matrix (outside the Sn-pipe) is highlighted using the dotted red line. (d) $NND_1$ and $NND_2$ of the Sn atoms located within the cube C1'. The $NND_k$ extracted from the APT maps are shown in colored histograms, while the corresponding distributions obtained from a randomized data set are shown using solid black lines. Inset: p-RDF analysis of the Sn atoms (relative to Sn) within the cube C1'. The p-RDF of unity is highlighted using the dotted black line.

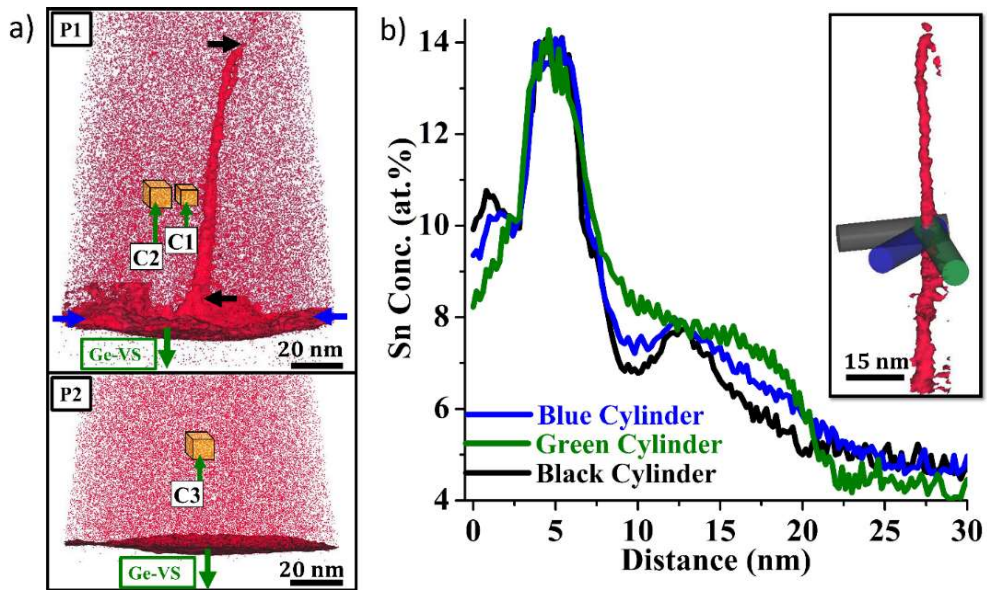

**Figure 2.** (a) Panel P1: The 3D atom-by-atom map (clipped from the entire reconstruction to highlight the region of interest) of the GeSn sample annealed at 320 °C, for 5 mins, showing a Sn-pipe. Only the Sn atoms (red dots) are shown for clarity. The Ge-VS is highlighted, and the interface of the BL with the Ge-VS is marked by double blue arrows. Similar to Figure 1(a), some Sn segregation can be seen to have occurred at the defective BL/Ge interface of the sample. C1 and C2 are two cubes placed proximally to the dislocation (C1 closer to the dislocation than C2). The dislocation is marked with double black arrows. Atoms located within the cubes were extracted out for statistical analysis (details in text). Panel P2: A second APT data set of the same sample, showing the BL and a part of the Ge-VS. A reference cube C3 at the BL is shown. (b) Sn concentration profile along the length of the black, the blue, and the green, coplanar cylinders that intersect the Sn-pipe. Inset: The placement of the three cylinders relative to the Sn-pipe. The Sn-pipe was identified by drawing iso-concentration surfaces within the APT map, at 6.9 at.% Sn concentrations (see the double black arrows in panel P1 in (a)).

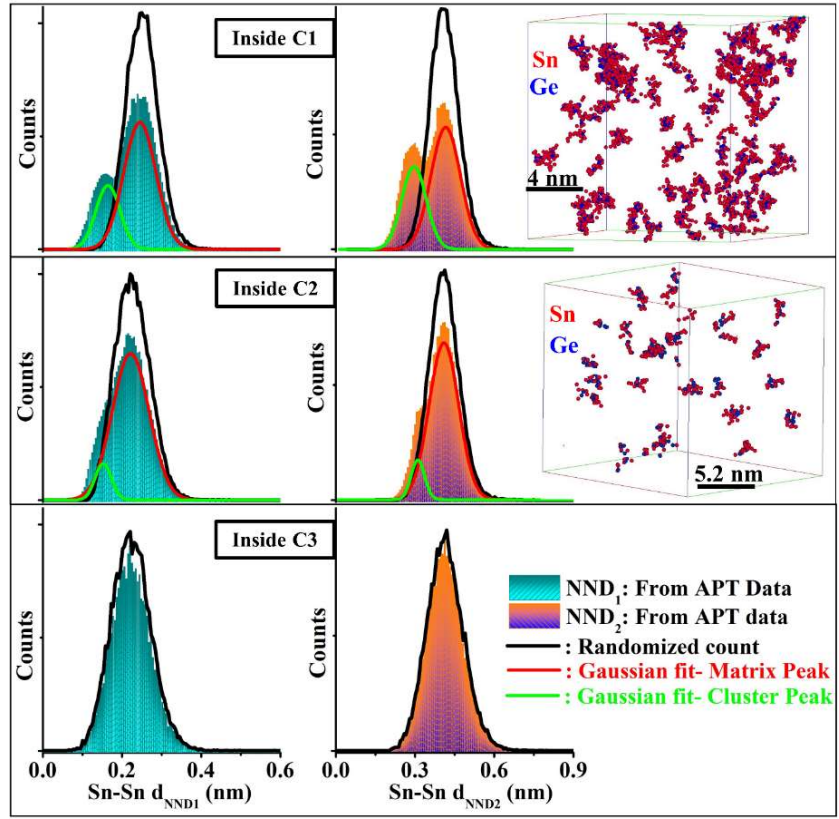

**Figure 3.** $NND_{1,2}$ analysis of the solute Sn atoms within the cubes C3 (bottom panel), C2 (middle panel), and C1 (top panel). Fig. 2 shows the location of the cubes within the sample annealed at 320 °C, for 5 min. In all the graphs, the $NND_k$ extracted from the APT maps are shown in colored histograms, while the corresponding distributions obtained from a randomized data set are shown using solid black lines. The bimodal nature of the experimental $NND_k$ in C2 and C1 are highlighted using two Gaussian fitting curves, shown used the solid red and green lines. The middle panel also shows the distribution of NRAC within C2. The top panel also shows the distribution of NRAC within C1.

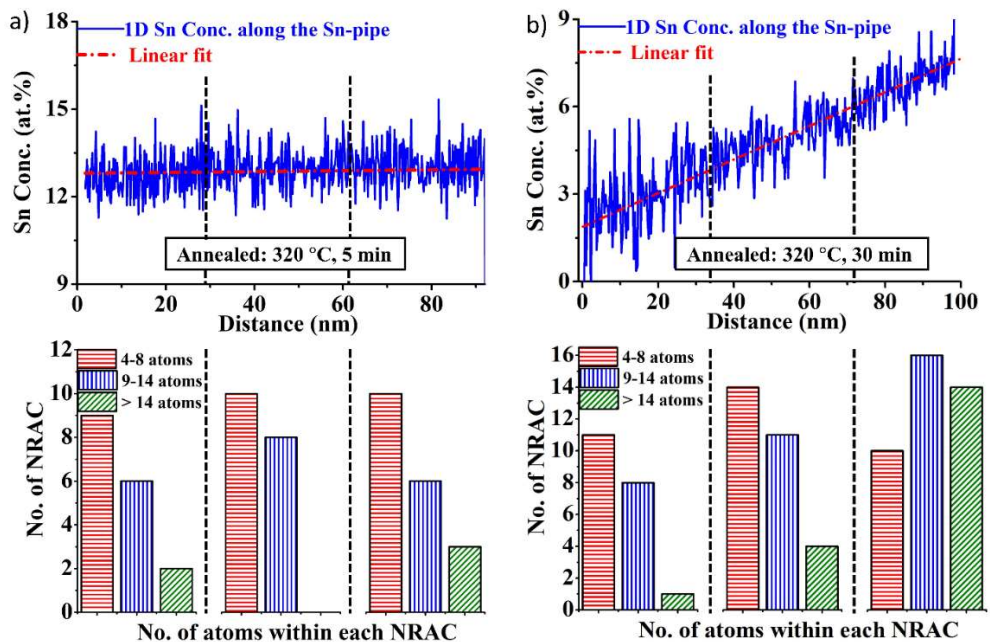

**Figure 4.** (a) Top Panel: Sn concentration profile within the Sn-pipe marked by the double black arrows in Figure 2(a): Panel 1 (sample annealed at 320 °C, 5 min). The Sn-pipe has been divided into approximately three equal sections (demarcated by the black dotted vertical lines). The cluster analysis was performed within each section, separately. Bottom Panel: Histograms showing the variation of the number of NRAC against the number of atoms within each NRAC, within each of the three sections of the Sn-pipe. (b) Top Panel: Sn concentration profile within the Sn-pipe marked by the double black arrows in Figure 1(a) (sample annealed at 320 °C, 30 min). Similar to (a), the Sn-pipe has been divided into approximately three equal sections (demarcated by the black dotted vertical lines) for subsequent cluster analysis. Bottom Panel: Histograms showing the variation of the number of NRAC against the number of atoms within each NRAC, within each of the three sections of the Sn-pipe. The liner fits of the concentration profiles in (a) and (b) are shown using the red dotted lines. Also, in (a) and (b), the top of the Sn pipes is to the right (larger x-values in 'nm' in the concentration profiles).

# Supplementary Information

# Atomic Pathways of Solute Segregation in the Vicinity of Nanoscale Defects

Samik Mukherjee,[†, *] Simone Assali,[†] Aashish Kumar,[†] and Oussama Moutanabbir[†, *]

[†]Department of Engineering Physics, École Polytechnique de Montréal, C. P. 6079, Succ. Centre-Ville, Montreal, Québec H3C 3A7, Canada

* samik.mukherjee@polymtl.ca
* oussama.moutanabbir@polymtl.ca

**Details of APT and HAADF-STEM measurements:** Before the APT tip fabrication in Dual-FIB, a 50 nm thick Ni capping layer was co-deposited on all the samples (using an electron-beam evaporator) to protect the top-most part of the samples from ion-implanted damage during the tip fabrication process. In this work, the field evaporation of individual atoms in the APT was assisted by focusing a picosecond pulsed UV laser ($\lambda = 355$nm), with a beam waist smaller than 5µm, on the apex of the needle-shaped specimen. The laser pulse repetition rate was maintained at 500 kHz throughout. The evaporation rate (ion/pulse) was varied between $0.2 – 2.0$ over a single run. The laser pulse energy was also varied between $3.0 – 25$ pJ over a single run. The base temperature and base pressure within the APT chamber was maintained at 30 K and $3.2\times10^{-11}$ Torr, respectively. The 3-D reconstructions were performed using Cameca's IVAS program.

The HR-STEM analysis was conducted in a double cross-section-corrected FEI Titan microscope, operated at 200 kV, using a convergence angle of 19.1 mrad. CEOS CESCOR corrector was used to yield a resolution of 0.8 Å. The images were recorded using a HAADF detector and the data was processed using the digital micrograph GMS3 software. For the STEM-EELS analysis, the spectrometer entrance aperture was set to 5 mm, giving a collection semi-angle of 55 mrad. The signal integration time was 5 ms. The spot size for the measurements was set to 9 nm. Aberration corrected magnetic lenses helped in making the probes of the order of 1-2 Å diameter with a beam current of 250 pA. A dispersion of 1eV/channel was used.

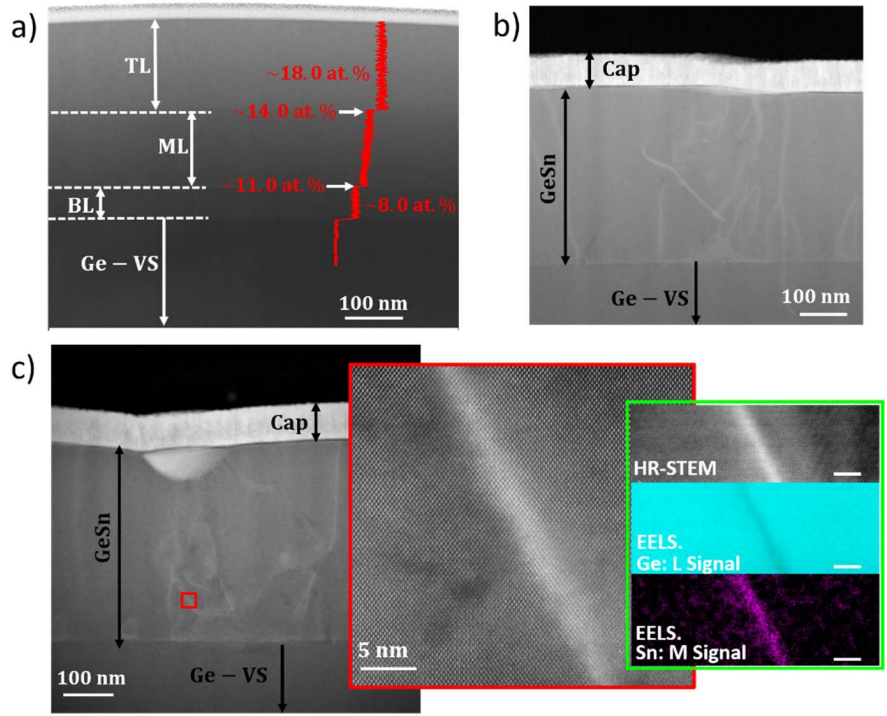

**Figure S1.** (a) HAADF-STEM image of the as-grown GeSn sample. The TL, ML, BL, and Ge-VS as well as the TL/ML, ML/BL, and BL/Ge-VS interfaces have been highlighted. The Sn concentration of the as-grown sample has been overlaid (in red) in the STEM image. (b) HAADF-STEM- images of GeSn sample annealed at 320 ℃ for 30 mins, showing the capping layer, the GeSn layer, and a part of the Ge-VS. (c) HAADF-STEM- images of GeSn sample annealed at 320 ℃ for 5 mins. Inset: High-resolution HAADF-STEM image of dislocation and the corresponding EELS maps of the Ge L signal, and the Sn M signal. The scale bars of the EELS maps is 5 nm.

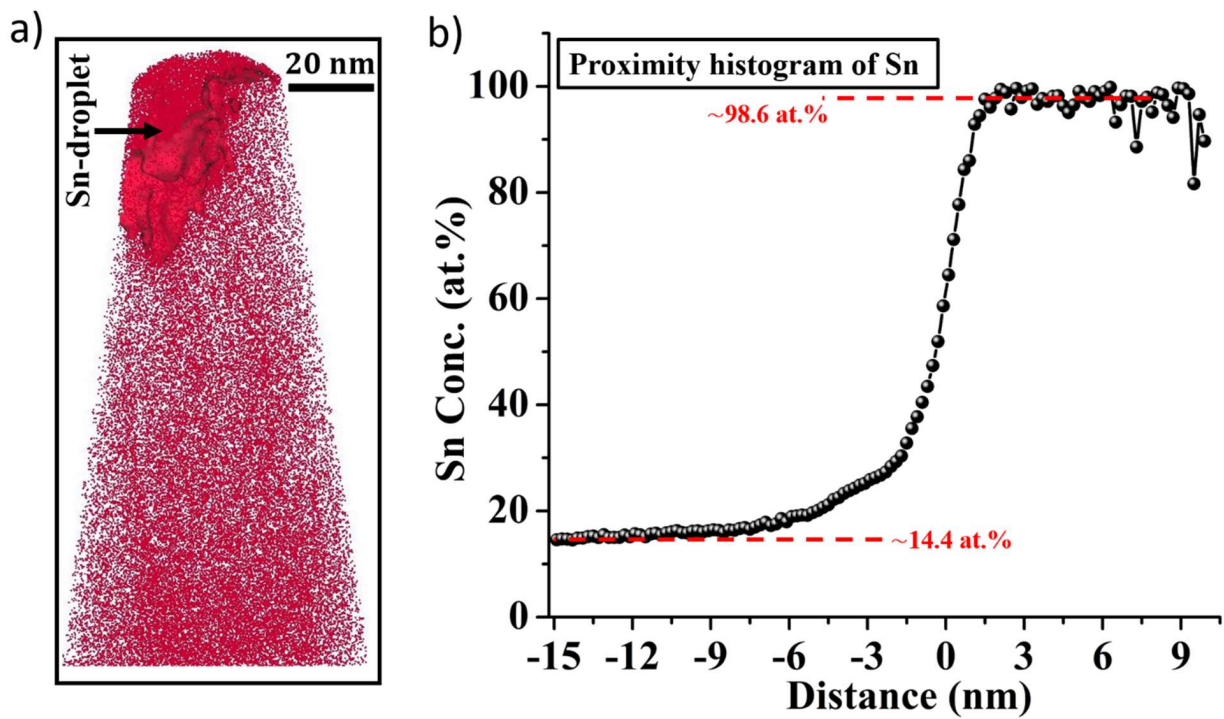

**Figure S2.** (a) The third APT dataset of the GeSn sample annealed at 320 °C for 5 mins, showing a solidified Sn droplet at the surface. The solidified droplet was identified by drawing iso-concentration surfaces at 41.5 at.% Sn concentrations. (b) Proximity histogram, binned at 0.2 nm, of the Sn atoms relative to the interfaces of solidified Sn droplet. The red dotted lines show the mean Sn concentration within the droplet (~98.6 at.%) and far from the droplet within the TL (~14.4 at.%).

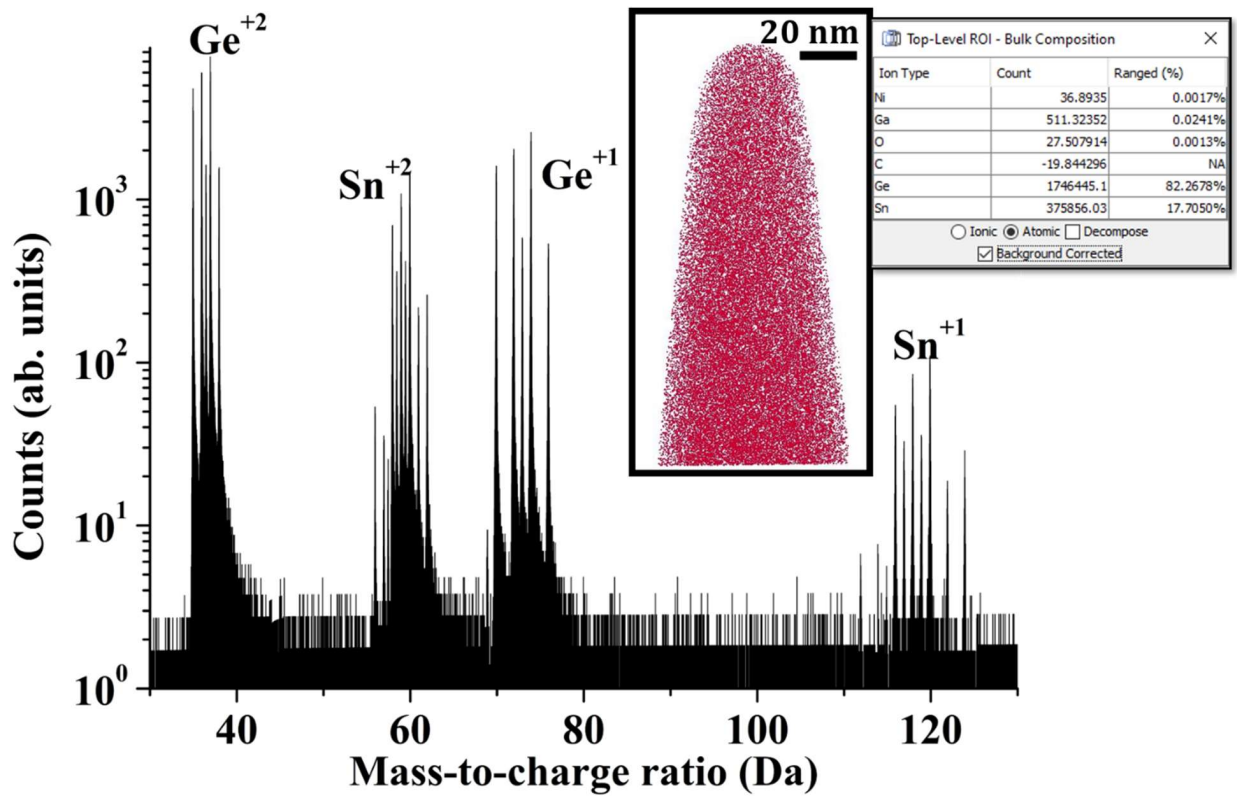

**Figure S3.** The mass-spectrum of the fourth APT data set of the GeSn sample, annealed at 320 °C for 5 mins. Inset: The corresponding 3D atomic distribution, showing the TL to be without any dislocation or solidified droplet. The bulk normalized Sn concentration (from mass-spectrum analysis) of the TL can be seen to be ~17.7 at.%.

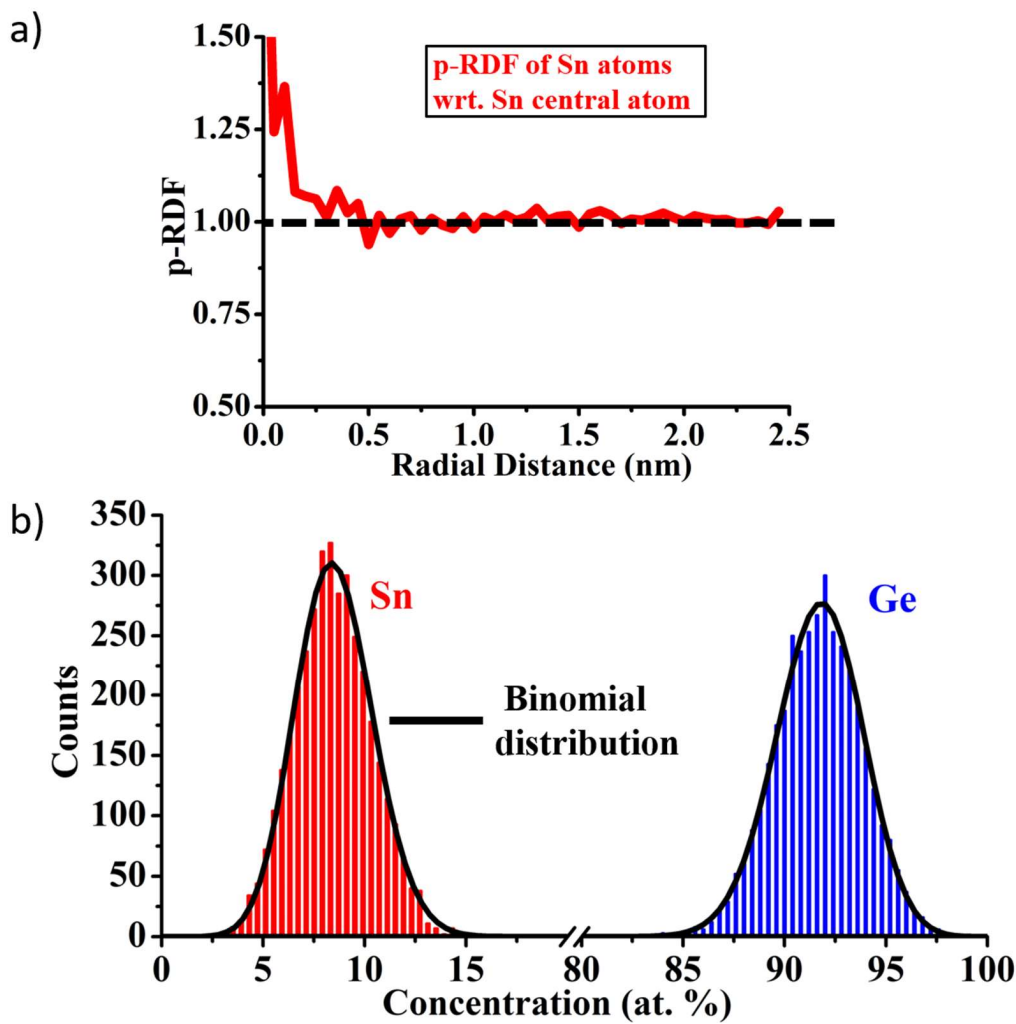

**Figure S4.** This figure pertains to the atoms within the cube C3 in panel P2 of Figure 2(a) of the main manuscript. (a) p-RDF analysis of the Sn atoms, relative to Sn. The p-RDF of unity is highlighted using the dotted black line. (b) The frequency distribution (in histograms) of Ge and Sn atoms (binned at 100 atoms), showing that the atomic distribution within C3 follows a binomial distribution (solid black line), with the distributions reaching peak values at the expected atomic concentrations of ~8.0 at.% for Sn and ~92.0 at.% for Ge. In the frequency distribution analysis, the region of interest is broken down into 'N' blocks each containing '$n_b$' atoms. The total number of atoms of a particular element is then counted in each block and the frequency at which a particular atom occurs is then compared with a binomial distribution. For example, if 'n' is the number of atoms of a particular element that is randomly distributed throughout the volume and has a bulk normalized concentration of 'C' then its frequency of occurrence must follow the binomial distribution $f(n) = \{Nn_b!/n!\,(n_b - n)!\}C^n(1 - C)^{(n_b - n)}$.